# From the Hands of an Early Adopter's Avatar to Virtual Junkyards: Analysis of Virtual Goods' Lifetime Survival

**Kamil Bortko** [1], **Patryk Pazura** [1], **Juho Hamari** [2,3], **Piotr Bartków** [1] and **Jarosław Jankowski** [1,*]

[1] Faculty of Computer Science and Information Technology, West Pomeranian University of Technology, Szczecin, Poland; kbortko@wi.zut.edu.pl (K.B.); patryk.pazura@gmail.com (P.P.); pbartkow@wi.zut.edu.pl (P.B.)

[2] Gamification Group, Faculty of Information Technology and Communication Sciences, Tampere University, 33100 Tampere, Finland; juho.hamari@tut.fi

[3] Gamification Group, Faculty of Humanities, University of Turku, 20500 Turku, Finland

* Correspondence: jjankowski@wi.zut.edu.pl

**Abstract:** One of the major questions in the study of economics, logistics, and business forecasting is the measurement and prediction of value creation, distribution, and lifetime in the form of goods. In "real" economies, a perfect model for the circulation of goods is impossible. However, virtual realities and economies pose a new frontier for the broad study of economics, since every good and transaction can be accurately tracked. Therefore, models that predict goods' circulation can be tested and confirmed before their introduction to "real life" and other scenarios. The present study is focused on the characteristics of early-stage adopters for virtual goods, and how they predict the lifespan of the goods. We employ machine learning and decision trees as the basis of our prediction models. Results provide evidence that the prediction of the lifespan of virtual objects is possible based just on data from early holders of those objects. Overall, communication and social activity are the main drivers for the effective propagation of virtual goods, and they are the most expected characteristics of early adopters.

## 1. Introduction

Virtual worlds and games have been postulated to provide unprecedented possibilities for research in general [1,2], but especially for the study of economics [3] due to their ability to systematically track every event in that reality, but also due to the possibility of creating controllable environments while having people exhibit natural behaviors.

Perhaps one of the most prominent veins of study related to virtual economies has been the study of consumer behavior related to adopting and purchasing virtual goods in virtual worlds and games [4–7]. This has especially been the case since games and virtual world operators have been the forerunners in implementing the so-called freemium or free-to-play business model ([8–10]), where playing or using the virtual environment is free of charge, but the operator generates revenue through different manifold marketing strategies combining classical sales tactics imbued with platform design that further encourages virtual-goods purchases [11–13].

Virtual goods mostly take up the forms of in-game items related to the theme of the game, such as avatar clothing, gear, vehicles, pets, emoticons, and other customization options [5,14], as well as different types of items related to the recent proliferation of "gamblification", where acquiring virtual goods is increasingly based on gambling-like mechanics, effectively blurring the line between gaming and gambling [15].

The largest vein of research in this continuum has been the investigation into why people purchase virtual goods [4,5] in primary or secondary markets within the virtual world. Popularly, this question was initially motivated by the sheer anecdotal amazement of why people would spend considerable amount of real money on products that "do not exist" [11,16]. However, since the initial combination of hype and disillusionment, virtual and game economies have entered into the realm of everyday consumer-facing services. Studying the question of why people purchase and trade virtual goods has primarily focused on latent psychological factors such as motivations, attitudes, experiences, and belief, and how they predict virtual-goods transactions as well as the internal design of the environment (see, for example, Reference [4] for a review of the area). However, the limitation within this sphere of research is that it can only provide a glimpse of the reasons why users purchase virtual goods as a singular event since it is focused on the consumer rather than the object of consumption and trade—the virtual good itself. Only few studies [17] have taken the initiative in an attempt to map the longer lifespan of virtual goods from their inception to circulation and to their ultimate end, destroyed from the virtual world, forgotten in a user's virtual bag, or existing in an account of a user who has stopped visiting the virtual world.

Additionally, one of the major hurdles in governing and maintaining virtual economies, in addition to increasing consumer demand for virtual goods [11], has been the balancing act between "sources" and "sinks" [18] of virtual goods within a virtual economy. There is no practical or technical reason why any virtual good could not exist in complete abundance within the virtual economy. However, this would create problems both in relation to the meaningfulness of acting within the virtual world due to extreme inflation, which would also effectively void any need for users to purchase or trade virtual goods. Therefore, the lifetime management of virtual goods is of vital importance for any virtual-economy operator (see References [6,11,18]). Some of the methods in the game-operator palette have been, for example, contrived durability and planned obsolescence of virtual goods (see, for example, Reference [19]).

Game developers are confronted with issues identified with the ideal recurrence of virtual-product updates, their volumes, and intensity, with an emphasis on ceaseless development [20]. Reduced recurrence of updates can result in user churn, while the consistent improvement of new content increases operational expenses. From another perspective, users may have a constrained capacity for digital content used when content is updated as often as possible. This might be regarded as unwise budget allocation when content production is fundamentally higher than demand. The life expectancy of web-based gaming items is generally shorter than that of traditional items, and users always expect system updates and new content [21,22]. Another issue is the habituation impact resulting from the short life expectancy of virtual products, and the limited time in which the item can attract online users. This opens up new research directions since, so far, it has principally been researched for traditional markets [23].

To address this research problem, the present study is focused on the characteristics of early-stage adopters of virtual goods and how they predict the lifespan of the goods. Rogers [24] treats 2.5% of users as innovators, 13.5% of users as early adopters, 34% as an early majority, and 34% and 16% as the late majority and laggards, respectively. This research shows how characteristics of early-stage adopters affect user engagement and product lifespan. The main contributions include the identification of the role of early adopters of virtual goods for product lifespan, and building a predictive model for product life with the use of data.

The empirical study is followed by analysis based on survival prediction models and identification of the role of the characteristics of early-stage adopters for product lifespan. Decision trees showed the ability to predict product lifespan with the use of product-adopter characteristics. The rest of the paper is organized as follows. The Methodology section contains the conceptual framework, dataset description, and methodological background. The Results section includes descriptive statistics and results from the lifespan models based on user characteristics. This is followed by results from product classification in terms of their lifespan and user characteristics with an accuracy higher than 80%. The study is concluded in the final section.

## 2. Methodology

### 2.1. Research Questions and Study Design

The presented study assumes the ability of virtual-product survival prediction with user attributes, especially those interested in the product at different stages of the product lifecycle. This research is based on the conceptual framework presented in Figure 1. A set of virtual products, Pi, was introduced to the audience of a social platform. Behaviors related to user engagement and products usage were collected. The node position within the example social network is represented by node size. Small circles were used for low degree nodes with one connection through medium sizes up to biggest ones for nodes with four connections. In general, user characteristics can represent various attributes related to network centrality and activity within the system like communication frequency and intensity of platform usage. They create parameters space with m distinguished variables assigned to each user in the form of vector $V = [V_1, V_2, ..., V_m]$. Users adopted to each product can be divided into five adoption groups with 2.5% of users interested in product distinguished as innovators, next 13.5% classified as early adopters, 34% as early majority, 34% of late majority, and users adopting to product at the end (laggards) as 16% of all adopters.

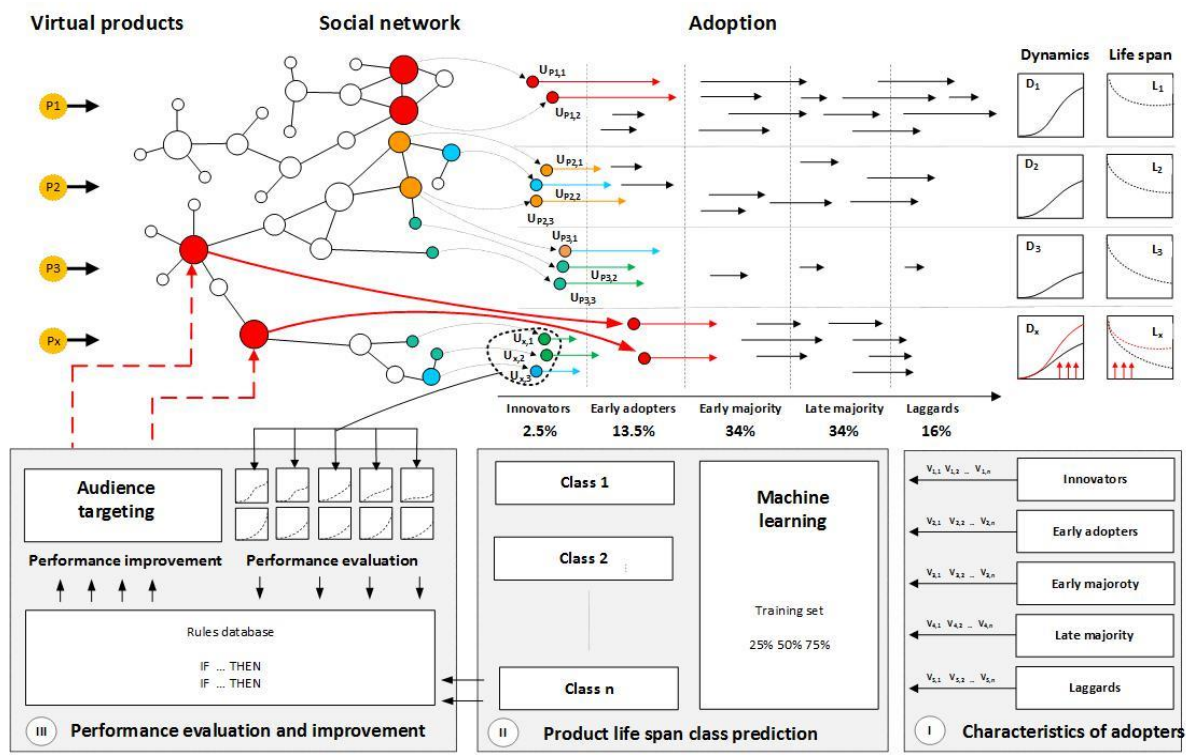

**Figure 1.** (**I**) Analytical system integration with the platform with the ability to detect the characteristics of users engaged in a new product, and the stages when adoption takes place; (**II**) product classification according to survival time and audience characteristics; (**III**) monitoring the performance of new products, predicting their usage, and additional audience targeting.

One of the research questions is whether innovator and early-adopter characteristics can affect product lifespan. It would be possible to identify the characteristics of initial users and, in cases of low-performance prediction, the interest of users with central network positions could be increased by sample delivery, trial accounts, or other incentives. As a result, followers and late adopters would be influenced and motivated for new-product usage.

Three exemplary possible scenarios are presented. In Scenario 1, Product P1 is introduced, and two users, namely, $UP_{1,1}$ and $UP_{1,2}$, with high positions represented by red nodes adopt the product as innovators. They are followed by users adopted at several stages of the product lifecycle, and it is considered a successful product launch, boosted by top-user engagement. The campaign is characterized by high dynamics $D_1$ and long product lifespan $L_1$. Scenario 2 for Product P2 assumes three innovators, $UP_{2,1}$, $UP_{2,2}$, and $UP_{2,3}$, characterized by medium metrics within a social network, and this is represented by orange and blue nodes. They build the interest of other users in the launched product, and overall campaign evaluation results in medium dynamics $D_2$ and product lifespan $L_2$. Scenario 3 assigned to Product P3 is based on the interest of innovators with the lowest network positions. It results in dynamics $D_3$ and lifespan $L_3$. In an analytical system, historical product data are used to analyze the influence of user characteristics, especially innovators and early adopters, on the product's lifespan and engagement among other users. This is based on three stages of data processing. In Stage (I), the characteristics of adopters from all groups are measured. In Stage (II), classification is performed to build class descriptors of users who are characteristic for a product with different survival time. Results are used to build a knowledge base and rules set for further use within the system and future product evaluation. In the next stage, new product Px is launched and introduced to the system. Innovators and early adopters were monitored, and prediction of the product lifespan was performed. If the product that is assigned to the class with possible low lifespan, actions to improve performance can be implemented by the selection of users with high network positions to build interest in the new product, denoted as red arrows. The main goal is to increase the dynamics of product consumption $D_x$ and its lifespan $L_x$. In practice, it can be performed by product samples, trial accounts, or various other forms of incentives.

2.2. Dataset Description and Participants

The experimental study is based on data from the virtual world and the use of avatars within the platform [25,26]. The introduced dataset covers information from 195 items included in the form of user avatars. Items are utilized in the virtual-world platform providing various forms of entertainment and chat functions. Graphical symbols represent users who all have the chance to participate in the life of the online network, with 850,000 accounts initiated. Clients interact in the space of public graphical rooms that are related to various themes. They can configure and supply their private rooms and also utilize web-based games and unique entertainment alternatives.

The fundamental functions of the service are related to chatting, meeting new people, communication, and creating social relations. Other features include clothes and virtual products, styles, avatars, and a decorative element. New-product information can be distributed through private messages, sent through the use of an internal communication system. The analytical module concentrated on new items and this enhanced monitoring of content distribution and collecting information related to data-dissemination procedures. Clients accessed various amounts of functions that are commonly available, and also paid for premium services, which provide more potential outcomes. Virtual products appear in the form of products equivalent to real goods, special effects for avatars, or avatars themselves. Account extensions used within the system had different characteristics and purposes. For example, animations, flashing elements, and active objects handled by avatars were used.

While innovation-diffusion theory emphasizes the role of innovation characteristics, it was important to take into account objects with similar characteristics to minimize the impact of individual product features and the level of innovation. This led to analysis of comparable static-avatar elements with similar characteristics without special effects usually attracting more attention than static objects.

2.3. Survival Analysis Methods for Measuring Product Lifespan

The presented study uses survival analysis to analyze the expected time duration when interest in new products exists, which represents the product lifespan. In the field of survival analysis length of time taken is referred to event time [27]—product usage time in our case. Survival analysis was originally developed in the medical field, as a means of analyzing the time between medical intervention and death. Over the past few decades, the field was expanded to include other events as well as events that occur multiple times for a given individual [28].

Survival analysis has wide applications in the field of marketing, including customer-relationship management (CRM), marketing-campaign management, and trigger-event management [29]. If we denote the time taken for an event to occur as T, we can construct a frequency histogram and model a series of events as a function of time. The probability distribution function for T can be denoted by $f(t)$. The cumulative distribution function can be denoted by $F(t)$. This provides the following equation:

$$F(t) = p(T \leq t)$$

Using the above approach, we can represent survival as a function of time $S(t)$ such that: for $t = 0$, $S(t) = 1$ for the specific time that a failure occurs, the value of $S(t)$ is zero [30]. In some cases, the time to failure will not be observable and only partial observation will be possible. In this case, we consider a specific 'censoring time' c. The survival function is then denoted as:

$$S(t) = P(T > t) = 1 - F(t)$$

Instantaneous hazard or conditional failure rate is the instantaneous rate at which a randomly selected individual—who is known to be alive at time (t 1) and will die at time t [31]. Mathematically, instantaneous hazard is equal to the number of failures between time t and time $t + D(t)$, divided by the size of the population at risk (at time t), divided by $D(t)$. This gives us the proportion of the present population at time t that fail, per unit of time, represented by the equation:

$$h(t) = \lim_{Dt \to 0} \frac{P(t < T \leq t + D(t) | T > t)}{D(t)} = \frac{f(t)}{S(t)}$$

Widely used, the Kaplan–Meier method is used to estimate time-related events [27]. Most commonly, it is used in biostatistics to analyze death as outcome. However, in more recent years, the technique has seen adoption in the fields of social sciences and industrial statistics. For example, in economics, we might measure how long people tend to remain unemployed after being let go by an employer; in engineering, we might measure how long a certain mechanical component tends to last before mechanical failure takes place. The survival function is theoretically a smooth curve, but it can be estimated using the Kaplan–Meier (KM) curve. Plotting the Kaplan–Meier estimate entails a series of horizontal steps of declining magnitude that, for a sufficiently large sample approach, estimate the true survival function for the given population. When applying this approach, survival-function value between successive sampled observations is presumed constant [32]. An important advantage of the Kaplan–Meier curve is its ability to take into account censored data loss within the sample before the final outcome is observed. In cases where no truncation or censoring occurs, the Kaplan–Meier curve is equivalent to empirical distribution [33,34].

As mentioned, survival analysis has wide applications for marketing, including CRM, marketing-campaign management, and trigger-event management [29]. Depending on the business setting, e.g., contractual versus noncontractual, different techniques can be applied [29]. For example, a goal might be to analyze the performance of a marketing campaign (while in progress), and how different customer features affect its performance. In this case, recurrent survival analysis techniques are used and the hazard function models the tendency of customers to buy a given product [35,36]

Survival analysis also has wide applications in the field of customer-behavior analysis. Among other things, it has been used to make predictions regarding customer retention in the banking [37] and insurance industries [38], credit scoring (with macroeconomic variables) [39], credit-granting decisions [40], and risk predictions of small-business loans [41].

Aside from customer behavior, survival analysis has been used to make predictions regarding the survival of online companies [42], as well as the duration of open-source projects [43]. Similarly, product survival in given markets was analyzed with network effects based on product compatibility [44].

The advent of digital marketing has provided additional streams of rich behavior data and subsequently new fertile ground for the application of survival analysis. With these data, survival analysis can be used to make predictions regarding the survival of music albums and distribution [45], the survival of mobile applications [46], as well as e-commerce recommendations to users [47].

For social platforms, survival analysis has been applied to triadic relationships within a social network [48], as well as participation in online entertainment communities with the use of entertainment and community-based mechanisms [49]. Player activity in online games provides valuable data for analysis, with a focus on game hours, subscription cancellations [50], and the adjustment of game parameters. In this context, a primary goal is to achieve the optimal user experience in terms of game speed and design [51].

Another area that is being explored is churn prediction in mobile games using survival ensembles [52] and player-motivation theories [53]. While game-time survival analysis can be used as a predictor of user engagement, it can also provide knowledge regarding factors that affect gameplay duration [54]. Similarly, it can provide insight in how player activity and popularity affects retention within games [55]. It can also be used to uncover predictors of game-session length, such as character level or age within the game [56]. The ability to quantify user satisfaction provides greater ability to target user needs [57].

2.4. Classification Methods Used for Product-Lifespan Prediction

Decision-making involves several approaches, including decision-tree classifiers [58]. Making a decision based on the structure of a decision tree allows complex decisions to be broken into a few small ones to deeply understand a problem. Decision trees are pervasive in a variety of real-world applications, including and not limited to medicinal research [59], biology, credit risk assessment, financial-market modeling, electrical engineering, quality control, biology, chemistry and so on. The evolution of web applications and social media resulted new areas of decision support and data analytics focused on user interaction and online behaviors. Decision trees are used for e-commerce, social media, online games, player segmentation, and other areas. Among other areas, applications include decision-tree usage for the future adoption of e-commerce-service predictions [60]. In social media, decision trees are used, for example, to predict the distance between users with Twitter activity data [61] and Twitter message classification with the use of the Classification and Regression Tree (CART) algorithm [62]. This wide area of applications includes online games with a focus on player-segmentation strategies based on self-recognition and game behaviors in the online game world to improve player satisfaction [63]. Integrated data-mining techniques such as association rule discovery, decision trees, and self-organizing map neural networks within the Kano model are used for customer-preference analysis in massively multiplayer online role-playing games [64].

Predicting aspects of playing behavior with the use of supervised learning algorithms is trained on large-scale player-behavior data. Decision-tree learning induces well-performing and informative solutions [65]. Rule databases can be used in a form of rule reasoner in online games for the detection of cheating activities [66], while a case-based reasoning approach can be applied for the purpose of training our system to learn and predict player strategies [67]. Educational games can be improved with decision trees used for the identification of factors affecting user behavior and knowledge acquisition within educational online games [68]. In other applications, decision trees are used for Internet game addiction in adolescents [69] and game-traffic analysis at the transport layer [70].

Clusterization techniques are used for player-behavior segmentation in computer games with the use of K-means and simplex volume maximization clustering [71], and user segmentation is used for retention management in online social games [72]. Integrated data-mining and experiential-marketing techniques can be used to segment online-game customers [73].

Owing to their structure, trees are easy to interpret, and hence result in better insights to problems. Nodes in decision-tree ramify from root nodes, and each node represents a condition related to a single input variable (feature), each branch represents a condition outcome, and each leaf node represents the class label. In this study, we applied CART [74], which is a binary tree. The method is to generate binary-tree-utilized binary-recursive partitioning that divides the dataset into two subsets, as per the minimization of a heterogeneity criterion computed on the resulting subsets. Each division made is based on a single variable, and some variables may not be used at all, while others may be used several times. Each subset is then further split based on independent rules.

Let's take into account decision tree T, with one of its leaves $t$. T is a mapping that assigns a leaf $t$ to each sample $(X_i^1, \ldots, X_i^p)$, where $i$ is an index for the samples. T can be viewed as a mapping to assign a value $\hat{Y}_i = T(X_i^1, \ldots, X_i^p)$ to each sample. Let $p(j|t)$ be the proportion of a class $j$ in a leaf $t$. The Gini index and entropy are the two most popular heterogeneity criteria. The entropy index is:

$$E_t = \sum_j p(j|t) \log p(j|t)$$

with, by convention, $x \log x = 0$ when $x = 0$. The Gini Index is an impurity-based criterion that measures divergence between the probability distributions of the target attribute's values [75]. The Gini index is defined as:

$$D_t = \sum_{i \neq j} p(i|t) p(j|t) = 1 - \sum_{i \neq j} p(i|t)^2$$

For the purpose of our research, we followed the formal definitions proposed by Maimon and Rokach [76], with bag algebra in the background [77]. Following the definitions, the training set in typical supervised learning consists of labeled examples in order to form a description that can be used to predict previously unseen examples. Many data descriptions were created, and the most frequently used is the bag instance of a certain bag schema. The bag schema is denoted as $R(A \cup y)$ and provides the description of the attributes and their domains. A indicates the set of input attributes containing n attributes: $A = \{a_1, \ldots, a_i, \ldots, a_n\}$ and y represents the class variable or the target attribute. Attributes appear in one of two forms, nominal or numeric. If attribute $a_i$ is nominal, we denote it by $dom(a_i) = \{v_{i,1}, \ldots v_{i,2}, \ldots, v_{i,|dom(a_i)|}\}$ where $dom(a_i)$ stands for its finite cardinality.

The domain of the target attribute appears in a similar way, $dom(y) = \{c_1, \ldots, c_{|dom(a_i)|}\}$. All possible examples that make up the set are called instance space: $X = dom(a_1) \times dom(a_2) \ldots \times dom(x_n)$. The Cartesian product of all input-attribute domains define the instance space.

The Cartesian product of all input-attribute domains and target-attribute domain defines the universal instance space, i.e., $U = X \times dom(y)$. Training consists of a set of tuples. Each tuple is described by a vector of attribute values. The training set is denoted as $S(R) = (\langle x_1, y_1 \rangle, \ldots, \langle x_n, y_n \rangle)$ where $x_q \in X$ and $y_q \in dom(y)$. The algorithm needs these data to learn how to match the input variables with the dependent variables—briefly, how to fit into the algorithm.

The test dataset was used to verify how our algorithm learns from the training data by checking its classification accuracy. We achieved this through matching classified observation with a real-observation class.

## 3. Results

3.1. Descriptive Statistics

Statistical analysis was based on 195 elements divided into four types of virtual elements, E1, E2, E3, and E4, used within the system representing avatar head, body, legs, and shoes. The data contain the anonymized behavioral patterns of 8139 unique users. The analyzed products were introduced to system users within 21 content updates (CUs).

In order to perform statistical analysis, we used two groups of separate variables related to user activities. Variable abbreviations and their explanation can be found in Table 1.

**Table 1.** Abbreviations of variables with their short description used in the article.

|  | Short | Explanation of the Variables |
|---|---|---|
| Activity Factors | CA | communication activity |
|  | SD | social dynamics |
|  | CP | communication popularity |
|  | SP | social position |
|  | AA | adoption activity |
| Experience Factors | FR_in | all messages sent by the user until they are changed to unique users |
|  | FR_out | all messages received by the user until changed from unique users |
|  | MSG_in | all messages sent by the user until the change |
|  | MSG_out | all messages received by the user until the change |
|  | FR_total | total amount of FR_in and FR_out |
|  | MSG_total | total amount of MSG_in and MSG_out |
|  | U_log | number of logins before the change |
| Adoption Group | AG1 | innovators |
|  | AG2 | innovators + early adopters |
|  | AG3 | innovators + early adopters+early majority |
|  | AG4 | innovators + early adopters+early majority+late majority |
|  | AG5 | innovators + early adopters+early majority+late majority+laggards |

The first group includes five variables treated as Activity Factors with the symbols CA–AA. These are, respectively: CA, communication activity represented by an average number of messages received by users adopting the product divided by the number of logins; SD, social dynamics, represented by an average of a number of friends of the product adopter divided by the number of logins; CP, communication popularity, represented by an average number of outgoing messages divided by incoming messages; SP, social position, represented by the average number of received messages divided by the number of incoming messages; and AA, adoption activity, represented by averaging the number of new avatar-element usages divided by the number of logins.

The second group of variables represents Experience Factors related to user activity since account creation, such as MSG_in, the average number of all messages received by the user until the avatar changes; MSG_out, the average number of all messages sent by the user until the change; MSG_total, the average number of total messages sent and received by the user; FR_in—the number of unique friends contacting the user until the avatar change; FR_out—the number of friends contacted before the avatar usage, and FR_total, the average total number of friends.

For each product, users were assigned to Adoption Groups in five classes: innovators, early adopters, early majority, late majority, and laggards, according to time of adoption.

For the purpose of determining the role of used variables, user-related factors were used for the statistical models of survival analysis. We took into account the User Activity and User Experience factors. Initial analysis showed that, for most products, survival time was shorter than one month, and only few of them reached nearly three months. To cover usage periods with more detail, five time periods were taken into account during analysis: one week, two weeks, one month, two months, and three months. One week as the shortest period makes it possible to analyze behavior each day of the week after product launch. Analyzing the statistical significance of predictors that influenced the lifetime dependent variable, we can see that mean CA and AA showed statistical significance of $p < 0.05$ for all periods. The CP variable, on the other hand, is one that has no effect and is not relevant in any given period. Separately analyzing each period, we can see that the periods of one month, two months, and three months showed the significance of the CA, SD, SP and AA variables. Wald's statistics with results presented in Table 2 showed the highest value with CA in the periods of two weeks month and two months. In the three-month period, Wald pointed to the significance of AA. The influence of predictors on the dependent variable over seven days showed significance in CA, SD, and AA. However, in the 14 day period, only two predictors, CA and AA, showed statistical significance, which affected the product's life expectancy. In the next step, Kaplan–Meier (Figures 2–6) survival probability charts for one month with division for user parameters, and the three user groups were analyzed. The diagrams show the emergence of a growing number of increasingly shorter episodes that, at the border, seek the real function of survival. Figure 6 shows a survival model without division into classes as a general model for divisional and nondivisional variables.

The next stages show statistical regression models with division into aggregated groups of adopters, i.e., AG1–AG5. Regarding the explanation of these classes, we can refer to Table 3. Regression analysis was divided into two groups of variables, and product life is a dependent variable. The first group of variables (predictors) include average variable values from CA to AA. The second group include experience-related variables, i.e., MSG_in, MSG_out, FR_in, and FR_out. In Table 3, we can see the statistical-significance parameter (p) and the strength-of-significance factor (f).

The first group of predictors for AG1 showed significance for CA and AA. Average predictor AA was characterized by the strongest impact. For AG2, the case was definitely different. Four of the five predictors, i.e., CA and CP to AA, were significant. The only predictor that did not have statistical significance was average predictor SD. The impact forces of the predictors, especially in SP, were characterized by a strong accent. For AG3 and AG4, regression analysis showed similar significance to AG2 also for four predictors, but in these cases, lack of predictor significance in relation to dependent variables was shown by the mean of AA variables. In both cases, the CA variable strongly affected G5 results, where things were quite different. Statistical significance was only demonstrated in three cases: CA, CP, and SP.

The second group of predictors that affect the dependent variable also showed variability. In AG1, one of the four predictors was statistically significant, namely, FR_out (0.03). The situation looked completely different for AG2. Here, we can clearly see the strength of joining two classes. Significance statistics showed a positive result for up to three predictors, i.e., MSG_out, FR_in, and FR_out. In the case of AG3, AG4, and AG5, statistical significance was shown by 100% of predictors from FR_out, being the one that acts the strongest on the dependent variable.

The next part of analysis was based on an intergroup comparison of user characteristics between products with different survival time. In order to compare individual lifecycles with the Activity and Experience factors, we used the Mann–Whitney U Test. Analysis was presented in four perspectives: analysis of individual user classes, analysis of aggregated user classes based on activity-factor analysis of individual user classes, and analysis of aggregated user classes based on Experience Factors. Periods that we compared with each other are visible in Table A1. By starting division-variable predictor analysis for innovators, we can see the lack of significance of parameters at the first comparison period. In the next two, we can see that predictor CA was significant, which indicates that the periods significantly differed from predictor CA. In the last pair of compared periods, predictors CA, SD, and SP showed the largest differences.

Statistics for innovators show us a tendency for the comparative period to be smaller, in this case, two to three months, so more predictors influenced the differences. Analyzing the four other user classes, we see the opposite relationship. Starting with early adopters, where the differences could be seen in the four predictors in the first two pairs, in the next two the number of differences decreased. In the cases of early-majority, late-majority, and laggard users, significance statistics that point to differences are slowly blurred, as in the case of laggards, where in the last group of period comparisons we see the lack of significance of the given predictor data, which indicates low differences. Based on the aggregated user classes, we can see that the first combination of innovators and early adopters positively affects predictor significance, and this indicates large differences for most of the analyzed pairs (from three to four strongly affecting differences). We can see that the shorter the comparison period is, the smaller the differences are, such as two months versus three months. By analyzing the statistics of nondivisional variables that also include four period pairs, we see that statistics for the innovators themselves did not show any significance. We can see statistical significance at subsequent classes. Analyzing the remaining classes together, we see that differences in individual periods clearly increase. So, for early adopters, when analyzing the last two pairs of periods, statistical significance was less than 0.05, which indicates an increase in differences. By analyzing the last group, laggards, we could see that, in each group of periods, differences are clear and quite significant in each of the period pairs being compared. The same applies to aggregated users. Here, we compare the first period and, only in the case of AG2, in 14 days versus three months, we see the lack of slight differences between predictors. Other groups indicate strong differences, as we can see in Tables A2 and A3.

**Table 2.** Survival analysis with five user variables divided into five periods with Wald statistics and statistical significance showed.

| Variables | 7 Days | | 14 Days | | 1 Month | | 2 Months | | 3 Months | |
|---|---|---|---|---|---|---|---|---|---|---|
| | Wald | p | Wald | p | Wald | p | Wald | p | Wald | p |
| CA | 24.90 | <0.01 | 28.90 | <0.01 | 20.16 | <0.01 | 23.73 | <0.01 | 21.61 | <0.01 |
| SD | 4.99 | 0.03 | 2.82 | 0.09 | 6.09 | <0.01 | 4.11 | 0.04 | 4.49 | 0.03 |
| CP | 0.51 | 0.47 | 0.08 | 0.78 | 2.00 | 0.16 | 1.27 | 0.26 | 1.50 | 0.22 |
| SP | 2.30 | 0.13 | 1.25 | 0.26 | 7.12 | 0.01 | 9.71 | <0.01 | 12.37 | <0.01 |
| AA | 6.09 | <0.01 | 6.18 | <0.01 | 20.06 | <0.01 | 22.45 | <0.01 | 26.58 | <0.01 |

**Table 3.** Results of regression analysis showing how Activity Factors and Experience Factors are affecting user assignment to adoption group.

| Variables | AG1 | | AG2 | | AG3 | | AG4 | | AG5 | |
|---|---|---|---|---|---|---|---|---|---|---|
| | f | p | f | p | f | p | f | p | f | p |
| CA | 5.07 | 0.03 | 4.19 | 0.04 | 36.33 | <0.01 | 29.48 | <0.01 | 16.95 | <0.01 |
| SD | 0.62 | 0.43 | 0.10 | 0.75 | 7.97 | 0.01 | 4.27 | 0.04 | 0.39 | 0.53 |
| CP | 0.21 | 0.65 | 9.10 | <0.01 | 5.74 | 0.02 | 5.36 | 0.02 | 8.25 | <0.01 |
| SP | 0.52 | 0.47 | 13.98 | <0.01 | 9.62 | <0.01 | 9.74 | <0.01 | 13.47 | <0.01 |
| AA | 9.78 | <0.01 | 9.97 | <0.01 | 0.01 | 0.94 | 0.23 | 0.63 | 0.05 | 0.82 |
| MSG_in | 0.77 | 0.38 | 2.99 | 0.09 | 14.24 | <0.01 | 7.18 | 0.01 | 4.44 | 0.04 |
| MSG_out | <0.01 | 0.94 | 6.07 | 0.01 | 18.38 | <0.01 | 9.51 | <0.01 | 7.67 | 0.01 |
| FR_in | 1.22 | 0.27 | 6.62 | 0.01 | 13.51 | <0.01 | 21.22 | <0.01 | 12.83 | <0.01 |
| FR_out | 4.49 | 0.04 | 15.67 | <0.01 | 30.57 | <0.01 | 36.94 | <0.01 | 29.11 | <0.01 |

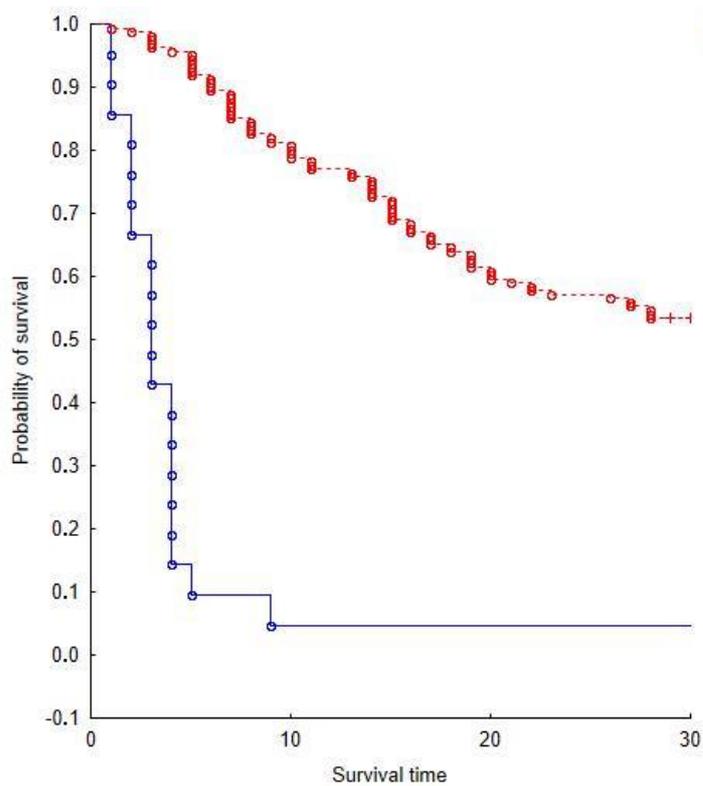

**Figure 2.** The Kaplan-Meier survival model for two groups of Experience Factors over a period of one month.

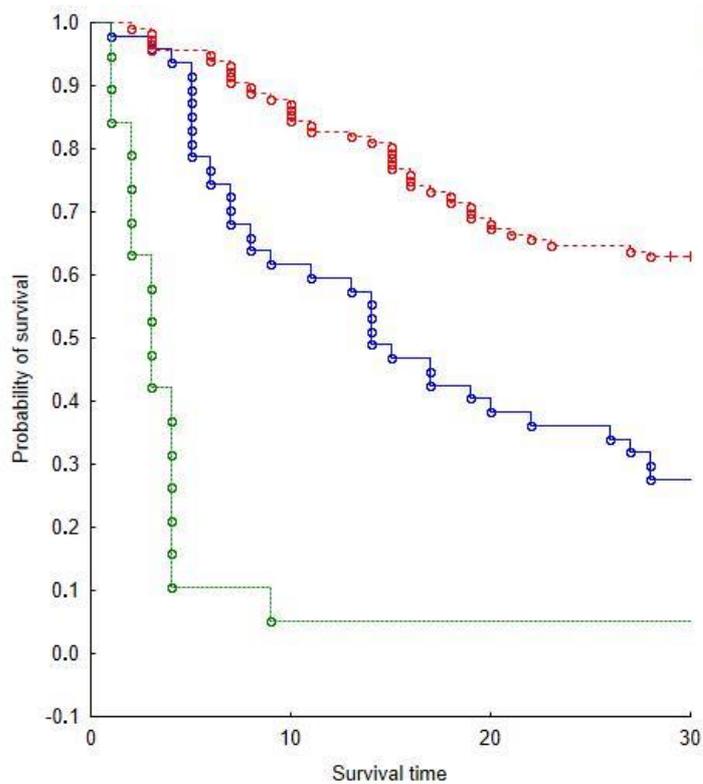

**Figure 3.** The Kaplan-Meier survival model for three groups of Experience Factors over a period of one month.

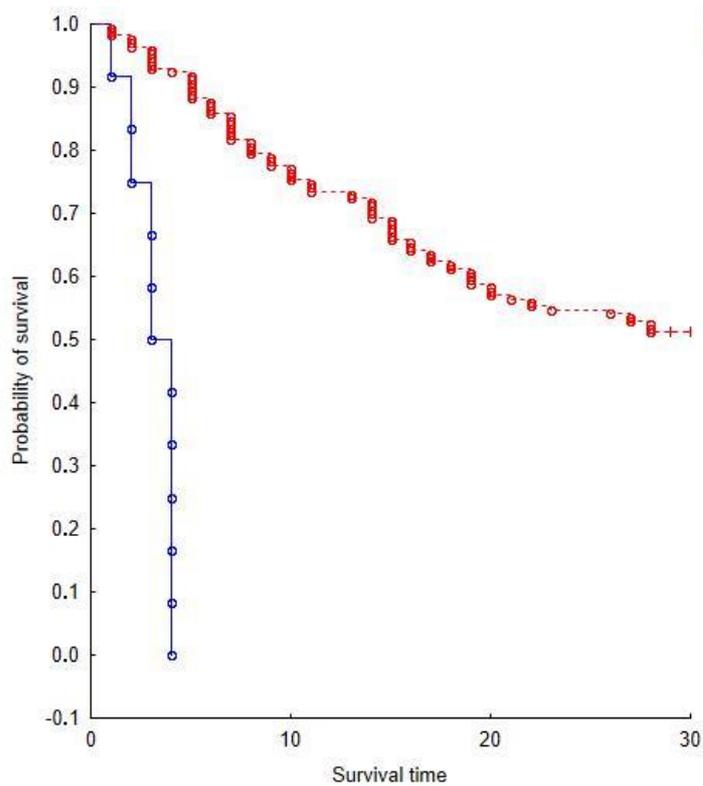

**Figure 4.** The Kaplan-Meier survival model for two groups of Activity Factors over a period of one month.

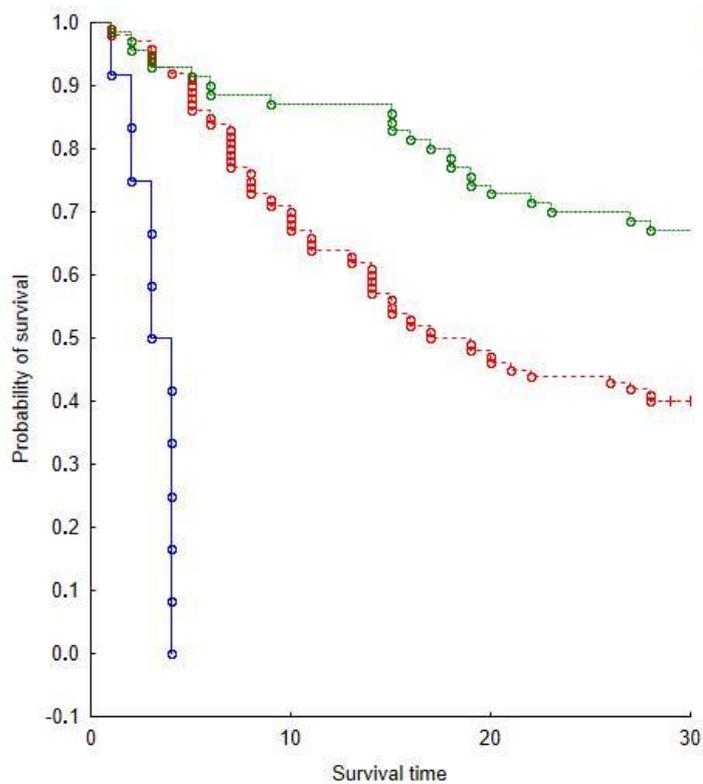

**Figure 5.** The Kaplan-Meier survival model for three groups of activity of experience over a period of one month.

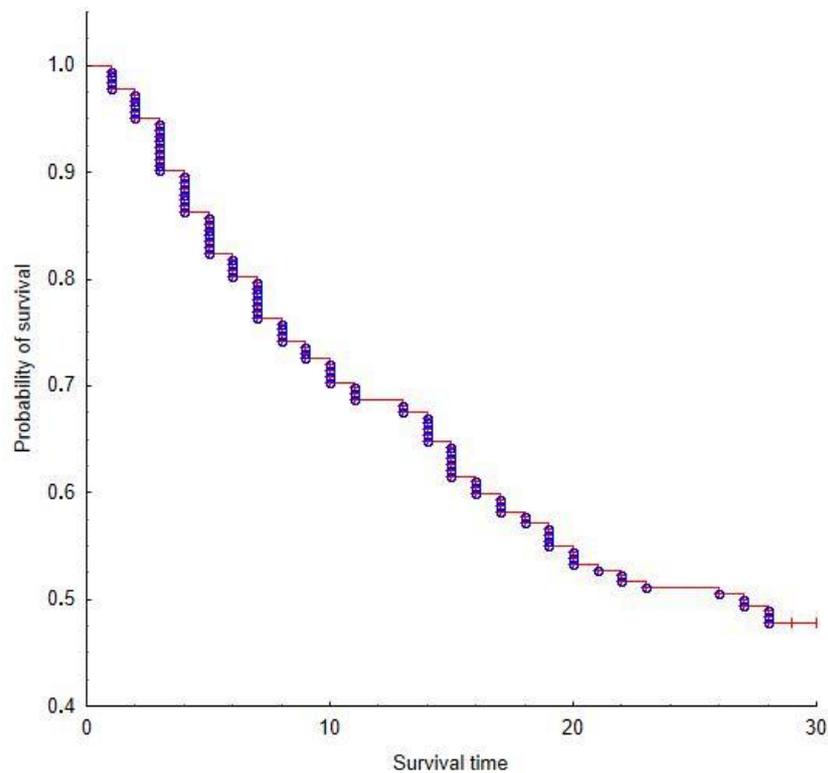

**Figure 6.** The Kaplan-Meier survival model for one group of divisional and non-divisional variables over a period of one month.

3.2. Survival-Time Prediction with Early-Adopter Characteristics

For survival-time, a prediction dataset containing the usage statistics of 195 newly introduced products was used. Usage statistics for each product are defined by product and user identifiers, and a timestamp representing time when the product (in this case, the avatar) was used by a specific user. For each product, only the first usage per user was taken into account. For each product, data were collected from a newly added product starting from product launch until last product usage. For each analysis, two sets of variables were used based on the User Activity and User Experience factors presented earlier in Table 1.

In Figure 7, we see a high increase in the CP variable for products with seven-day survival with simultaneous small CA values. In other periods, we see density with slight deviations, as in the case of the three-month period, where we see growth in the CA variable; in a 14-day survival period, an increase in the SD variable was observed. As in the previous chart, Figure 8 shows a clear division into survival-period groups. Within seven days, an increase in the AA ratio with a simultaneous drop in SD was visible, which may indicate a drop in interest from users with low SD. In the remaining periods, we can see in Figure 9 a clear decrease in the SD index with a simultaneous increase in CA; this showed that the more users communicate with others, the less likely it is for the product to be accepted.

In the case of this chart, we can clearly see that the fewer users log in, fewer messages are sent to others, and fewer sent to the circle of potential friends. There is a clear decline from that period to the next. In the case of the last graph, Figure 10, we can see density against the FR_out indicator at initial values oscillating at 150–250. Here, however, we also see a decline from period to period. In the initial period, the MSG_in indicator is small, but increases with survival time. However, the last period (three months) oscillates near the first period, which indicates a lower number of messages sent by the users adopting products in that group. Results from all Experience Factors and Activity Factors are presented in Figure A1 and Figure A2 within Supplementary Information.

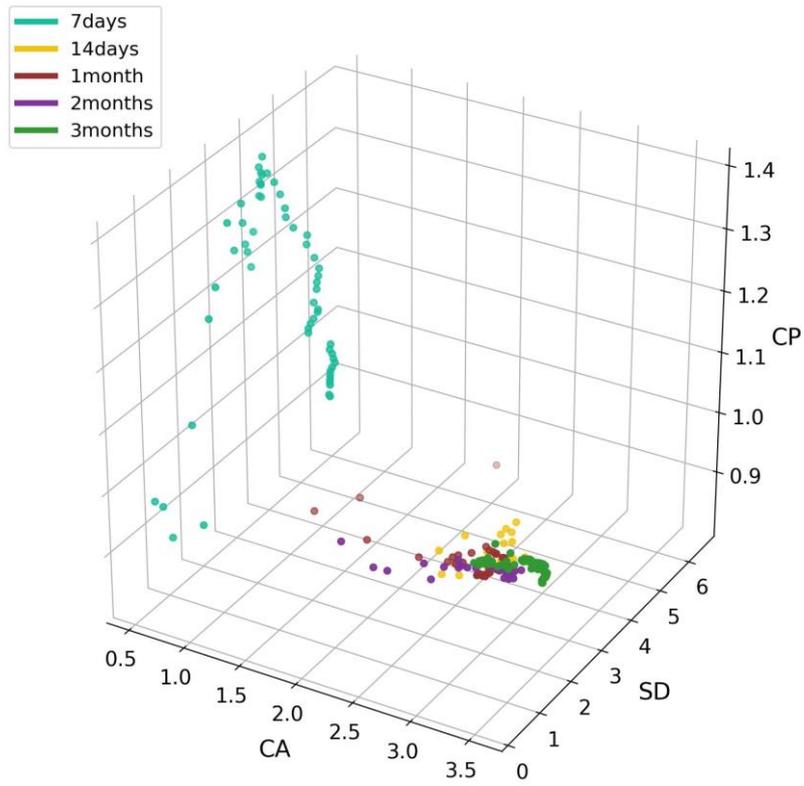

**Figure 7.** Dependence of objects in classes from CA, SD and CP variables.

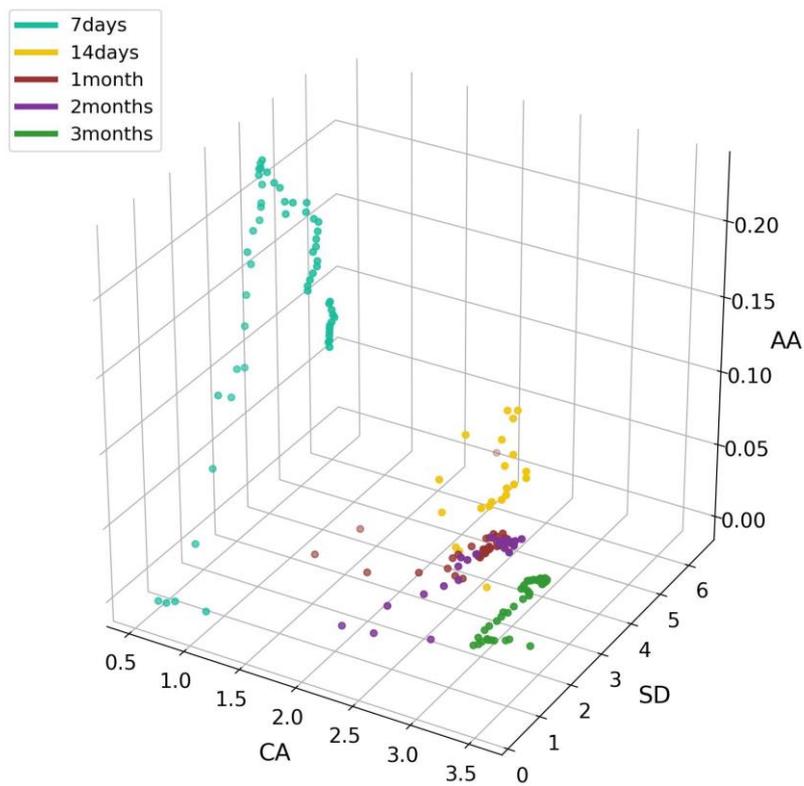

**Figure 8.** Dependence of objects in classes from CA, SD and AA variables.

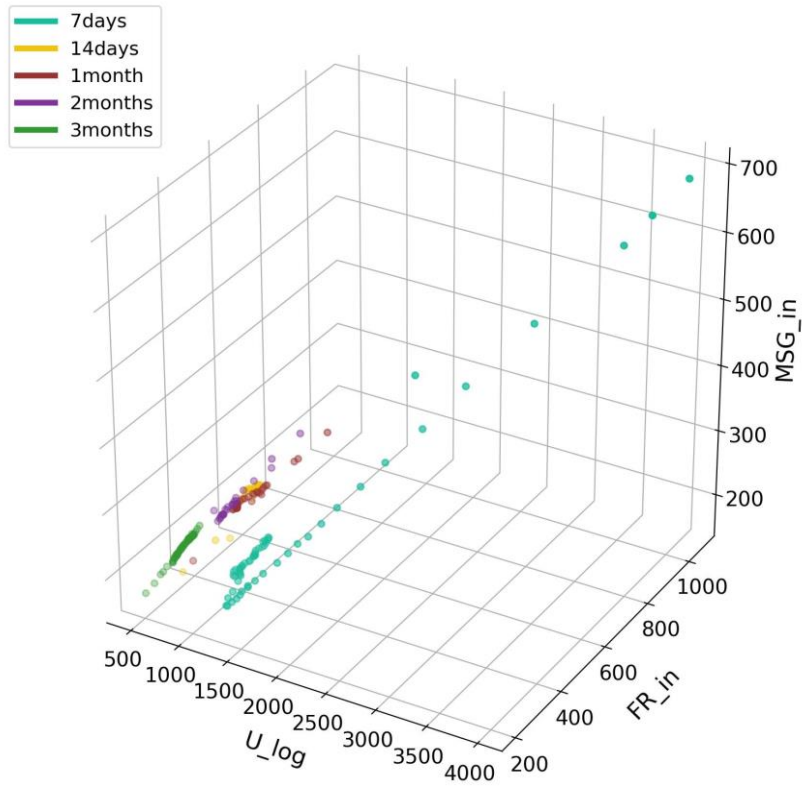

**Figure 9.** Dependence of objects in classes from U_log, FR_in and MSG_in variables.

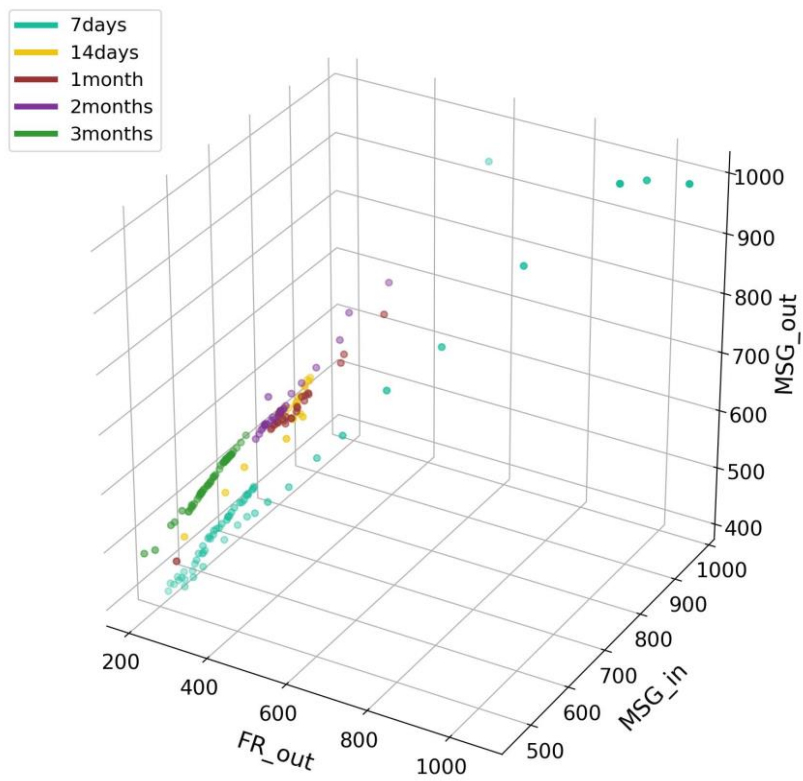

**Figure 10.** Dependence of objects in classes from FR_out, MSG_in and MSG_out variables.

Another stage investigates how the number of analyzed adopted users from 1% to 100% and their characteristics affect classification accuracy for the prediction of product lifetime and survival-class assignment (one week, two weeks, one month, two months, three months). The selection of observations to the training dataset was randomly performed; therefore, to stabilize the results, we repeated and averaged classification one hundred times for each dataset measure to obtain accurate information.

The experiment was carried out in three training-dataset sizes: 25%, 50%, and 75%. Classification and the decision-tree model were implemented with the help of the scikit-learn machine-learning library for the programming language Python. Classification was performed and, in the first stage, user-activity factors were used. Results are presented in Figure 11. They show high classification accuracy achieved for the training set based on 50% and 75% of the analyzed products. Accuracy at a level higher than 90% is achieved with less than 20% of product-usage statistics with activity factors taken into account. The training set based on 25% of the products delivered low accuracy, with a percentage of adopters lower than 60%, but it reached 90% when 70% of data were used for each product. Higher fluctuation of results was observed with a low number of analyzed adopted users.

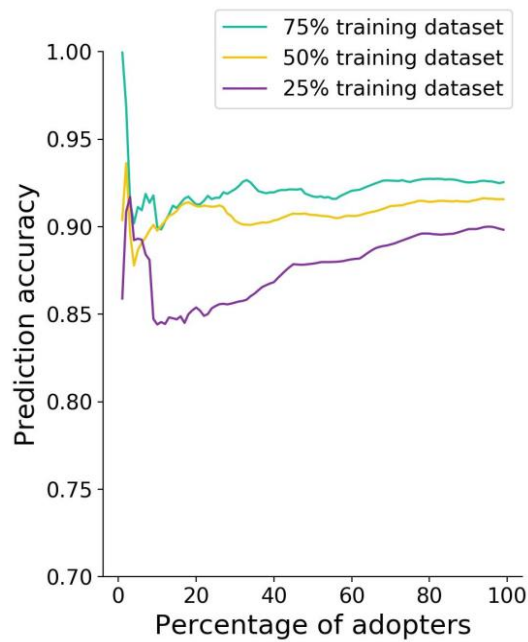

**Figure 11.** Accuracy of classification results with the use of Activity Factors for 25%, 50%, 75% training set and 10%, 20%, ..., 90% of adopters used.

Detailed numerical results are presented in Table 4. It shows that analysis of characteristics of even only 10% of product adopters makes it possible to predict product assignment to a class with low or longer survival time.

**Table 4.** Accuracy of classification results with the use of Activity Factors for 25%, 50%, 75% training set and 10%, 20%, . . . , 90% of adopters used.

| Traning Set | Number of Adopters Used for Classification | | | | | | | | |
|---|---|---|---|---|---|---|---|---|---|
| | 10% | 20% | 30% | 40% | 50% | 60% | 70% | 80% | 90% |
| 75 % | 0.900 | 0.915 | 0.920 | 0.920 | 0.917 | 0.921 | 0.926 | 0.927 | 0.925 |
| 50 % | 0.898 | 0.912 | 0.903 | 0.904 | 0.907 | 0.906 | 0.912 | 0.914 | 0.914 |
| 25 % | 0.844 | 0.854 | 0.857 | 0.868 | 0.879 | 0.881 | 0.890 | 0.896 | 0.899 |

Apart from social activity, factor classification was performed with the use of incremental data about user activity within the system. Results are presented in Figure 12. It shows that, for 50 and 70 training, the initial accuracy of the used low-fraction data at 1–3% of used data is very high due to innovator characteristics.

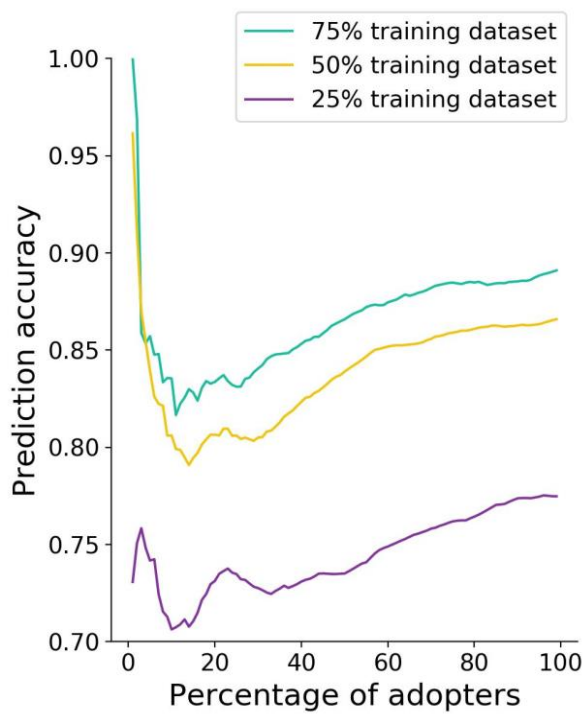

**Figure 12.** Accuracy of classification results with the use of Experience Factors for 25%, 50%, 75% training set and 10%, 20%, ..., 90% of adopters used.

Additionally used data dropped accuracy to 80%. Subsequently, it grew with data acquisition. For the training set with the size of 75% of used products, the lowest accuracy was achieved after using data from 15% of adopters, and 65% accuracy for 50% of products with 15% of the adopter sample used. For all training-set sizes, accuracy continuously grew together with the increased number of adopters.

Detailed numerical results for classification based on incremental usage statistics represented by Experience Factors for each user are presented in Table 5.

Table 6 shows classification-accuracy statistics with identified user groups as innovators, then innovators together with adopters, and extended by early majority, late majority, and laggards. For Activity Factors, this shows that even using data from only innovators (2.5% of first adopters) creates the ability to assign a product to one of five adoption classes. Innovators used together with adopters delivered results above 19% for training sets with 50% and 75% size. Classification based on the 25% training dataset delivered accuracy above 18%. Further connection of the adopter group slightly improved classification, but from a practical-application perspective, it delays the time during which product survival abilities are predicted and additional adopter targeting is performed. The worst results were obtained for Experience Factors, but they were still above 80% accuracy for the training sets with 50% and 75%.

**Table 5.** Accuracy of classification results with the use of Experience Factors for 25%, 50%, 75% training set and 10%, 20%, . . . , 90% of adopters used.

| Training Set | Number of Adopters Used for Classification | | | | | | | | |
|---|---|---|---|---|---|---|---|---|---|
| | 10% | 20% | 30% | 40% | 50% | 60% | 70% | 80% | 90% |
| 75% | 0.835 | 0.834 | 0.841 | 0.853 | 0.866 | 0.875 | 0.882 | 0.885 | 0.885 |
| 50% | 0.806 | 0.807 | 0.805 | 0.823 | 0.839 | 0.852 | 0.856 | 0.861 | 0.863 |
| 25% | 0.706 | 0.731 | 0.728 | 0.731 | 0.735 | 0.749 | 0.758 | 0.764 | 0.773 |

**Table 6.** Accuracy of classification results with the use of Activity and Experience Factors for combined adoption group.

| Group ID | Rogers Title | Size of Training Dataset | | | | | |
|---|---|---|---|---|---|---|---|
| | | Activity Factors | | | Experience Factors | | |
| | | 75% | 50% | 25% | 75% | 50% | 25% |
| G1 | innovators | 0.961 | 0.912 | 0.895 | 0.943 | 0.914 | 0.747 |
| G2 | + early adopters | 0.919 | 0.901 | 0.869 | 0.855 | 0.831 | 0.725 |
| G3 | + early majority | 0.919 | 0.905 | 0.865 | 0.849 | 0.821 | 0.729 |
| G4 | + late majority | 0.921 | 0.907 | 0.874 | 0.861 | 0.834 | 0.739 |
| G5 | + laggards | 0.922 | 0.908 | 0.878 | 0.865 | 0.839 | 0.744 |

## 4. Discussion and Conclusions

For expanded virtual product usage within online systems, new analytical models and strategies are required. Common phenomena to offline markets are regularly seen in electronic systems and are identified with lifespan, customer habituation, and new-product improvement techniques. This research indicates how the attributes of early adopters to new items can influence user engagement and the survival of virtual goods within dynamic electronic environments. Achieved results, from product classification based on decision trees, showed that it is possible to predict product lifespan with the use of adopter characteristics. Adopter communication activity, represented by Activity Factors, positively affected product survival time. This shows that adopters with high experience factors are the main influencers in the system, and their behavior is adopted by other users.

Monitoring of product-usage patterns and adopter characteristics makes it possible to identify products with possible low survival time, and invite additional adopters with the use of incentives and other techniques. Gathered knowledge can be used to reduce the habituation effect and increase product-usage time due to social influence and follower behavior.

Results from the conducted study lead to the following main conclusions:

characteristics of early adopters related to social activity positively influence product lifespan and the engagement of other users within the system;
product lifespan can be estimated with the use of initial-audience and early-adopter characteristics;
the combination of innovators and adopters positively affects the statistical significance of the dependent variable that represents survival time;
initial-user characteristics can be used to classify products in terms of future usage for the detection of low-potential products, for performance improvement and targeting additional adopters with the desired specifics.

Future work will concentrate on a progressive point-by-point evaluation of distribution in the use of social networks and behavior prediction, which is dependent on interpersonal organizations, and the use of conduct forecast, which is dependent on earlier behaviors.

**Author Contributions:** conceptualization, K.B., P.P., J.H., P.B. and J.J.; methodology, K.B., P.P., J.H., P.B. and J.J.; validation, K.B., P.P., P.B.; investigation, K.B., P.P. and P.B.; resources, P.B. and P.P.; data curation, P.B. and P.P.; writing–original draft preparation, K.B., P.P., J.H., P.B. and J.J.; writing–review and editing, K.B., P.P., J.H., P.B. and J.J.; visualization, K.B., P.P.; supervision, J.J.; project administration, J.J.; funding acquisition, J.J.

**Funding:** The work was supported by the National Science Centre of Poland, the decision no. 2017/27/B/HS4/01216.

**Conflicts of Interest:** The authors declare no conflict of interest.

## Appendix A

Supporting information

**Table A1.** Intergroup comparisons using the life expectancy of a product as a dependent variable and as divisional predictors divided into two groups.

| Variables | | Activity Factors | | | | | | | |
|---|---|---|---|---|---|---|---|---|---|
| | | 7 Days vs. 3 m | | 14 Days vs. 3 m | | 1 m vs. 3 m | | 2 m vs. 3 m | |
| | | p | z | p | z | p | z | p | z |
| innovator | CA | 0.21 | 1.25 | 2.00 | 0.05 | <0.01 | 3.17 | <0.01 | 2.88 |
| | SD | 0.28 | 1.07 | 0.41 | 0.69 | 0.27 | 1.10 | 0.04 | 2.10 |
| | CP | 0.64 | 0.47 | 0.69 | 0.49 | 0.32 | 0.99 | 0.74 | 0.33 |
| | SP | 0.11 | 1.62 | 0.42 | 0.67 | 0.08 | 1.75 | 0.02 | 2.26 |
| | AA | 0.85 | 0.19 | 1.29 | 0.20 | 0.93 | 0.09 | 0.36 | 0.91 |
| early adopter | CA | <0.01 | 5.90 | 5.57 | 0.00 | <0.01 | 2.84 | 0.01 | 2.64 |
| | SD | <0.01 | 3.17 | 4.22 | <0.01 | <0.01 | 4.14 | <0.01 | 3.99 |
| | CP | 0.52 | 0.64 | 1.64 | 0.10 | 0.44 | 0.77 | 0.57 | 0.57 |
| | SP | <0.01 | 4.50 | 4.46 | <0.01 | <0.01 | 2.85 | 0.06 | 1.88 |
| | AA | <0.01 | 3.70 | 2.39 | 0.02 | 0.15 | 1.44 | 0.55 | 0.60 |
| early majority | CA | <0.01 | 8.13 | 5.61 | <0.01 | <0.01 | 2.96 | 0.50 | 0.67 |
| | SD | <0.01 | 5.14 | 5.48 | <0.01 | <0.01 | 4.48 | 0.21 | 1.25 |
| | CP | 0.15 | 1.44 | 0.92 | 0.36 | 0.56 | -0.58 | 0.64 | 0.46 |
| | SP | 0.03 | 2.16 | 3.39 | <0.01 | 0.01 | 2.46 | 0.19 | 1.31 |
| | AA | 0.66 | 0.45 | 0.69 | 0.49 | 0.64 | 0.47 | 0.12 | 1.54 |
| late majority | CA | <0.01 | 7.00 | 1.37 | 0.17 | 0.05 | 1.94 | <0.01 | 3.17 |
| | SD | <0.01 | 3.17 | 2.41 | 0.02 | <0.01 | 2.87 | 0.40 | 0.84 |
| | CP | 0.04 | 2.04 | 1.65 | 0.10 | 0.68 | 0.41 | 0.79 | 0.27 |
| | SP | 0.06 | 1.86 | 1.20 | 0.23 | 0.65 | 0.46 | 0.03 | 2.18 |
| | AA | 0.71 | 0.37 | 2.30 | 0.02 | 0.76 | 0.31 | 0.36 | 0.92 |
| laggards | CA | <0.01 | 5.32 | 0.66 | 0.51 | 0.03 | 2.21 | 0.82 | 0.23 |
| | SD | 0.02 | 2.43 | 0.37 | 0.71 | 0.56 | 0.58 | 0.80 | 0.25 |
| | CP | 0.89 | 0.14 | 0.34 | 0.74 | 0.97 | 0.04 | 0.76 | 0.30 |
| | SP | 0.50 | 0.68 | 0.60 | 0.55 | 0.04 | 2.10 | 0.46 | 0.74 |
| | AA | 0.35 | 0.93 | 2.11 | 0.03 | 0.45 | 0.76 | 0.22 | 1.24 |
| All together | CA | <0.01 | 8.20 | 3.46 | <0.01 | 0.01 | 2.46 | 0.14 | 1.47 |
| | SD | 0.33 | 0.98 | 2.32 | 0.02 | <0.01 | 2.99 | 0.66 | 0.44 |
| | CP | 0.07 | 1.84 | 0.12 | 0.91 | 0.69 | 0.40 | 0.57 | 0.57 |
| | SP | <0.01 | 5.87 | 3.25 | <0.01 | 0.01 | 2.47 | 0.03 | 2.19 |
| | AA | <0.01 | 4.24 | 0.87 | 0.38 | 0.16 | 1.40 | 0.02 | 2.39 |
| G1 | CA | 0.21 | 1.25 | 2.00 | 0.05 | <0.01 | 3.17 | <0.01 | 2.88 |
| | SD | 0.28 | 1.07 | 0.41 | 0.69 | 0.27 | 1.10 | 0.04 | 2.10 |
| | CP | 0.64 | 0.47 | 0.69 | 0.49 | 0.32 | 0.99 | 0.74 | 0.33 |
| | SP | 0.11 | 1.62 | 0.42 | 0.67 | 0.08 | 1.75 | 0.02 | 2.26 |
| | AA | 0.85 | 0.19 | 1.29 | 0.20 | 0.93 | 0.09 | 0.36 | 0.91 |

Table A1. Cont.

|  | Variables | Activity Factors | | | | | | | |
|---|---|---|---|---|---|---|---|---|---|
|  |  | 7 Days vs. 3 m | | 14 Days vs. 3 m | | 1 m vs. 3 m | | 2 m vs. 3 m | |
|  |  | p | z | p | z | p | z | p | z |
| G2 | CA | <0.01 | 5.72 | 5.07 | <0.01 | <0.01 | 2.84 | <0.01 | 2.87 |
|  | SD | <0.01 | 2.92 | 3.59 | <0.01 | <0.01 | -3.43 | <0.01 | 3.80 |
|  | CP | 0.73 | 0.34 | 1.23 | 0.22 | 0.73 | 0.34 | 0.60 | 0.53 |
|  | SP | <0.01 | 4.02 | 4.36 | <0.01 | 0.01 | 2.68 | 0.01 | 2.47 |
|  | AA | <0.01 | 3.12 | 2.95 | <0.01 | 0.17 | 1.37 | 0.54 | 0.62 |
| G3 | CA | <0.01 | 8.20 | 5.63 | <0.01 | <0.01 | 3.25 | 0.03 | 2.11 |
|  | SD | 0.01 | 2.53 | 5.23 | <0.01 | <0.01 | 4.16 | 0.03 | 2.19 |
|  | CP | 0.70 | 0.38 | 0.08 | 0.93 | 0.48 | 0.71 | 0.64 | 0.47 |
|  | SP | <0.01 | 6.16 | 4.13 | <0.01 | 0.01 | 2.67 | 0.01 | 2.75 |
|  | AA | <0.01 | 3.14 | 0.11 | 0.92 | 0.86 | 0.17 | 0.91 | 0.12 |
| G4 | CA | <0.01 | 8.26 | 3.96 | <0.01 | 0.01 | 2.53 | <0.01 | 3.10 |
|  | SD | 0.12 | 1.56 | 3.29 | <0.01 | <0.01 | 3.37 | 0.62 | 0.49 |
|  | CP | 0.41 | 0.82 | 0.53 | 0.60 | 0.15 | 1.46 | 0.30 | 1.04 |
|  | SP | <0.01 | 5.80 | 3.55 | <0.01 | 0.01 | 2.45 | <0.01 | 3.04 |
|  | AA | <0.01 | 2.98 | 0.35 | 0.73 | 0.95 | 0.06 | 0.56 | 0.58 |
| G5 | CA | <0.01 | 8.20 | 3.46 | <0.01 | 0.01 | 2.46 | 0.14 | 1.47 |
|  | SD | 0.33 | 0.98 | 2.32 | 0.02 | <0.01 | 2.99 | 0.66 | 0.44 |
|  | CP | 0.07 | 1.84 | 0.12 | 0.91 | 0.69 | 0.40 | 0.57 | 0.57 |
|  | SP | <0.01 | 5.87 | 3.25 | <0.01 | 0.01 | 2.47 | 0.03 | 2.19 |
|  | AA | <0.01 | 4.24 | 0.87 | 0.38 | 0.16 | 1.40 | 0.02 | 2.39 |

**Table A2.** Intergroup comparisons using the life-length of a product as a dependent variable and as non-variable predictors.

|  | Variables | Experience Factors | | | | | | | |
|---|---|---|---|---|---|---|---|---|---|
|  |  | 7 Days vs. 3 Months | | 14 Days vs. 3 Months | | 1 Month vs. 3 Months | | 2 Months vs. 3 Months | |
|  |  | p | z | p | z | p | z | p | z |
| G1 | MSG_in | 0.99 | 0.01 | 0.16 | 0.88 | 0.38 | 0.87 | 0.93 | 0.09 |
|  | MSG_out | 0.13 | 1.53 | 0.48 | 0.63 | 0.55 | 0.6 | 0.81 | 0.24 |
|  | MSG_total | 0.83 | 0.21 | 0.13 | 0.90 | 0.47 | 0.72 | 0.87 | 0.17 |
|  | FR_in | 0.46 | 0.73 | 1.17 | 0.24 | 0.32 | 0.99 | 0.24 | 1.17 |
|  | FR_out | 0.21 | 1.25 | 1.61 | 0.11 | 0.91 | 0.11 | 0.10 | 1.67 |
|  | FR_total | 0.43 | 0.79 | 1.34 | 0.18 | 0.74 | 0.33 | 0.10 | 1.67 |
| G2 | MSG_in | 0.25 | 1.15 | 0.25 | 0.80 | 0.01 | 2.49 | 0.04 | 2.02 |
|  | MSG_out | 0.18 | 1.33 | 0.67 | 0.50 | 0.01 | 2.69 | 0.02 | 2.37 |
|  | MSG_total | 0.41 | 0.83 | 0.42 | 0.68 | 0.01 | 2.71 | 0.03 | 2.20 |
|  | FR_in | 0.01 | 2.6 | 0.25 | 0.80 | 0.03 | 2.15 | 0.01 | 2.47 |
|  | FR_out | 0.06 | 1.9 | 1.33 | 0.18 | <0.01 | 2.86 | <0.01 | 3.85 |
|  | FR_total | 0.03 | 2.12 | 0.82 | 0.41 | 0.01 | 2.64 | <0.01 | 3.44 |
| G3 | MSG_in | 0.39 | 0.86 | 3.92 | <0.01 | <0.01 | 3.61 | <0.01 | 3.12 |
|  | MSG_out | 0.69 | 0.4 | 4.09 | <0.01 | <0.01 | 3.91 | <0.01 | 3.27 |
|  | MSG_total | 0.96 | 0.05 | 4.05 | <0.01 | <0.01 | 3.79 | <0.01 | 3.22 |
|  | FR_in | 0.47 | 0.72 | 4.39 | <0.01 | <0.01 | 3.82 | <0.01 | 3.34 |
|  | FR_out | 0.92 | 0.10 | 4.90 | <0.01 | <0.01 | 4.76 | <0.01 | 4.24 |
|  | FR_total | 0.84 | 0.20 | 4.84 | <0.01 | <0.01 | 4.57 | <0.01 | 3.91 |
| G4 | MSG_in | 0.82 | 0.23 | 3.90 | <0.01 | 0.01 | 2.66 | <0.01 | 3.15 |
|  | MSG_out | 0.96 | 0.05 | 4.08 | <0.01 | <0.01 | 2.98 | <0.01 | 3.24 |
|  | MSG_total | 0.95 | 0.06 | 3.98 | <0.01 | <0.01 | 2.86 | <0.01 | 3.24 |
|  | FR_in | 0.29 | 1.07 | 3.93 | <0.01 | <0.01 | 2.99 | <0.01 | 3.62 |
|  | FR_out | 0.80 | 0.25 | 4.57 | <0.01 | <0.01 | 4.18 | <0.01 | 4.20 |
|  | FR_total | 0.90 | 0.12 | 4.38 | <0.01 | <0.01 | 3.83 | <0.01 | 4.01 |
| G5 | MSG_in | 0.18 | 1.33 | 3.57 | <0.01 | <0.01 | 4.44 | <0.01 | 3.36 |
|  | MSG_out | 0.4 | 0.85 | 3.72 | <0.01 | <0.01 | 4.47 | <0.01 | 3.43 |
|  | MSG_total | 0.34 | 0.96 | 3.65 | <0.01 | <0.01 | 4.48 | <0.01 | 3.43 |
|  | FR_in | 0.63 | 0.48 | 3.87 | <0.01 | <0.01 | 4.54 | <0.01 | 3.55 |
|  | FR_out | 0.16 | 1.40 | 4.31 | <0.01 | <0.01 | 4.98 | <0.01 | 3.97 |
|  | FR_total | 0.27 | 1.10 | 4.18 | <0.01 | <0.01 | 4.84 | <0.01 | 3.86 |

21 of 28**Table A3.** Intergroup comparisons using the life-length of a product as a dependent variable and as non-variable predictors.

|  | Variables | \multicolumn{8}{c}{Experience Factors} |
|---|---|---|---|---|---|---|---|---|---|
|  |  | 7 Days vs. 3 Months | | 14 Days vs. 3 Months | | 1 Month vs. 3 Months | | 2 Months vs. 3 Months | |
|  |  | p | z | p | z | p | z | p | z |
| innovator | MSG_in | 0.99 | 0.01 | 0.16 | 0.88 | 0.38 | 0.87 | 0.93 | 0.09 |
|  | MSG_out | 0.13 | 1.53 | 0.48 | 0.63 | 0.55 | 0.60 | 0.81 | 0.24 |
|  | MSG_total | 0.83 | 0.21 | 0.13 | 0.90 | 0.47 | 0.72 | 0.87 | 0.17 |
|  | FR_in | 0.46 | 0.73 | 1.17 | 0.24 | 0.32 | 0.99 | 0.24 | 1.17 |
|  | FR_out | 0.21 | 1.25 | 1.61 | 0.11 | 0.91 | 0.11 | 0.10 | 1.67 |
|  | FR_total | 0.43 | 0.79 | 1.34 | 0.18 | 0.74 | 0.33 | 0.10 | 1.67 |
| early adopter | MSG_in | 0.21 | 1.27 | 0.92 | 0.36 | <0.01 | 4.50 | <0.01 | 3.14 |
|  | MSG_out | 0.16 | 1.40 | 1.36 | 0.17 | <0.01 | 4.64 | <0.01 | 3.17 |
|  | MSG_total | 0.33 | 0.97 | 1.10 | 0.27 | <0.01 | 4.69 | <0.01 | 3.14 |
|  | FR_in | 0.02 | 2.42 | 0.29 | 0.77 | <0.01 | 3.18 | 0.02 | 2.34 |
|  | FR_out | 0.17 | 1.37 | 2.10 | 0.04 | <0.01 | 3.83 | <0.01 | 3.42 |
|  | FR_total | 0.12 | 1.56 | 1.37 | 0.17 | <0.01 | 3.71 | <0.01 | 3.10 |
| early majority | MSG_in | <0.01 | 3.96 | 5.34 | <0.01 | <0.01 | 4.66 | <0.01 | 3.31 |
|  | MSG_out | <0.01 | 2.93 | 5.37 | <0.01 | <0.01 | 4.73 | <0.01 | 3.31 |
|  | MSG_total | <0.01 | 3.66 | 5.31 | <0.01 | <0.01 | 4.75 | <0.01 | 3.38 |
|  | FR_in | 0.12 | 1.55 | 5.35 | <0.01 | <0.01 | 4.47 | <0.01 | 3.19 |
|  | FR_out | 0.38 | 0.88 | 5.48 | <0.01 | <0.01 | 4.89 | <0.01 | 3.18 |
|  | FR_total | 0.18 | 1.35 | 5.43 | <0.01 | <0.01 | 4.77 | <0.01 | 3.14 |
| late majority | MSG_in | 0.06 | 1.87 | 4.26 | <0.01 | <0.01 | 3.18 | <0.01 | 3.26 |
|  | MSG_out | 0.01 | 2.81 | 4.32 | <0.01 | <0.01 | 3.30 | <0.01 | 3.39 |
|  | MSG_total | 0.02 | 2.37 | 4.30 | <0.01 | <0.01 | 3.29 | <0.01 | 3.30 |
|  | FR_in | 0.81 | 0.24 | 3.80 | <0.01 | 0.01 | 2.59 | <0.01 | 3.07 |
|  | FR_out | 0.37 | 0.90 | 4.00 | <0.01 | <0.01 | 3.15 | <0.01 | 3.75 |
|  | FR_total | 0.53 | 0.63 | 3.97 | <0.01 | <0.01 | 3.02 | <0.01 | 3.43 |
| laggards | MSG_in | 0.03 | 2.19 | 3.15 | <0.01 | <0.01 | 3.16 | <0.01 | 3.32 |
|  | MSG_out | 0.01 | 2.64 | 3.36 | <0.01 | <0.01 | 3.29 | <0.01 | 3.37 |
|  | MSG_total | 0.02 | 2.36 | 3.31 | <0.01 | <0.01 | 3.29 | <0.01 | 3.41 |
|  | FR_in | 0.04 | 2.08 | 2.69 | 0.01 | <0.01 | 2.99 | <0.01 | 3.09 |
|  | FR_out | 0.04 | 2.01 | 2.30 | 0.02 | <0.01 | 2.89 | <0.01 | 2.87 |
|  | FR_total | 0.04 | 2.05 | 2.39 | 0.02 | <0.01 | 3.04 | <0.01 | 2.91 |
| All together | MSG_in | 0.18 | 1.33 | 3.57 | <0.01 | <0.01 | 4.44 | <0.01 | 3.36 |
|  | MSG_out | 0.40 | 0.85 | 3.72 | <0.01 | <0.01 | 4.47 | <0.01 | 3.43 |
|  | MSG_total | 0.34 | 0.96 | 3.65 | <0.01 | <0.01 | 4.48 | <0.01 | 3.43 |
|  | FR_in | 0.63 | 0.48 | 3.87 | <0.01 | <0.01 | 4.54 | <0.01 | 3.55 |
|  | FR_out | 0.16 | 1.40 | 4.31 | <0.01 | <0.01 | 4.98 | <0.01 | 3.97 |
|  | FR_total | 0.27 | 1.10 | 4.18 | <0.01 | <0.01 | 4.84 | <0.01 | 3.86 |

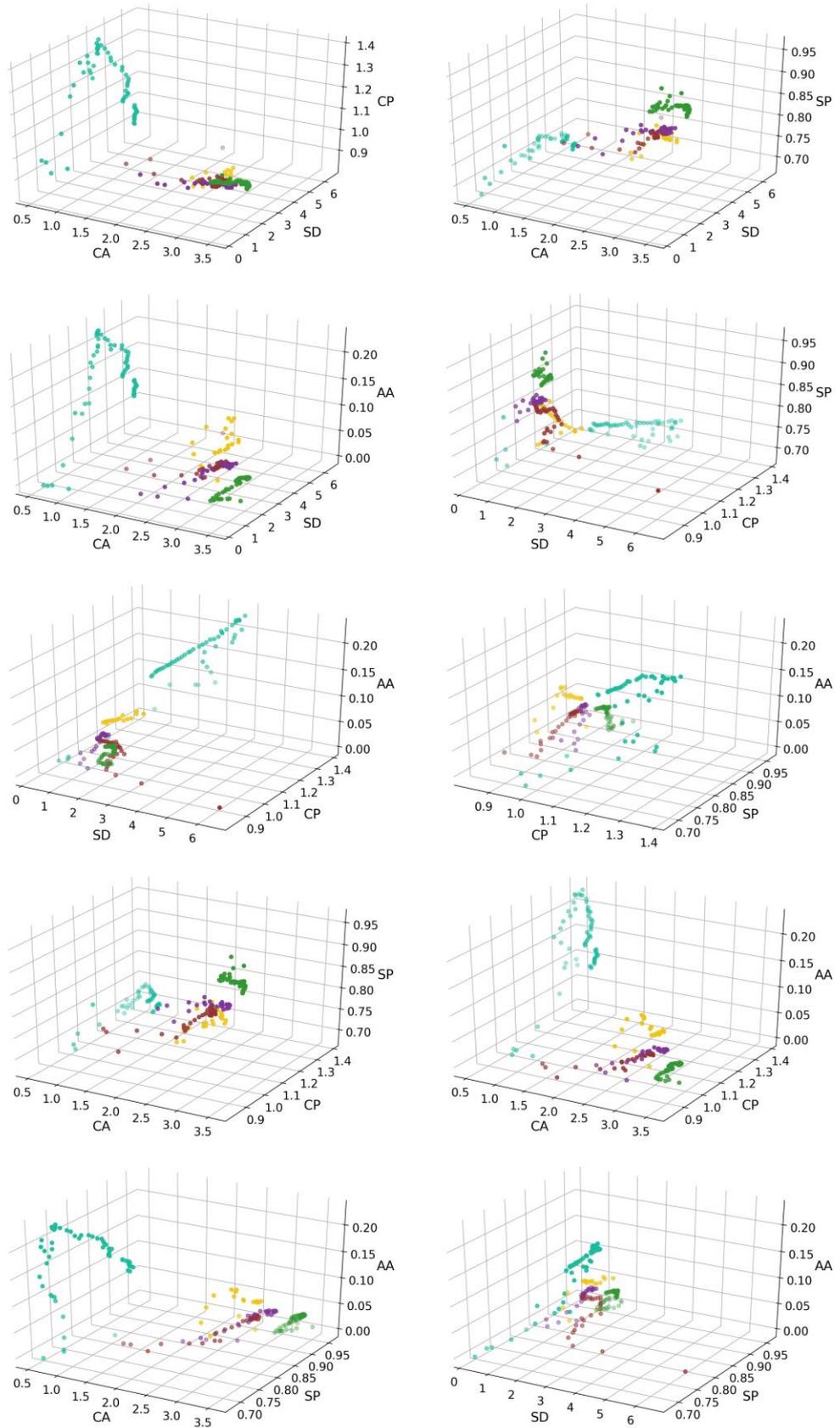

**Figure A1.** Dependence of objects in classes from all Activity Factors.

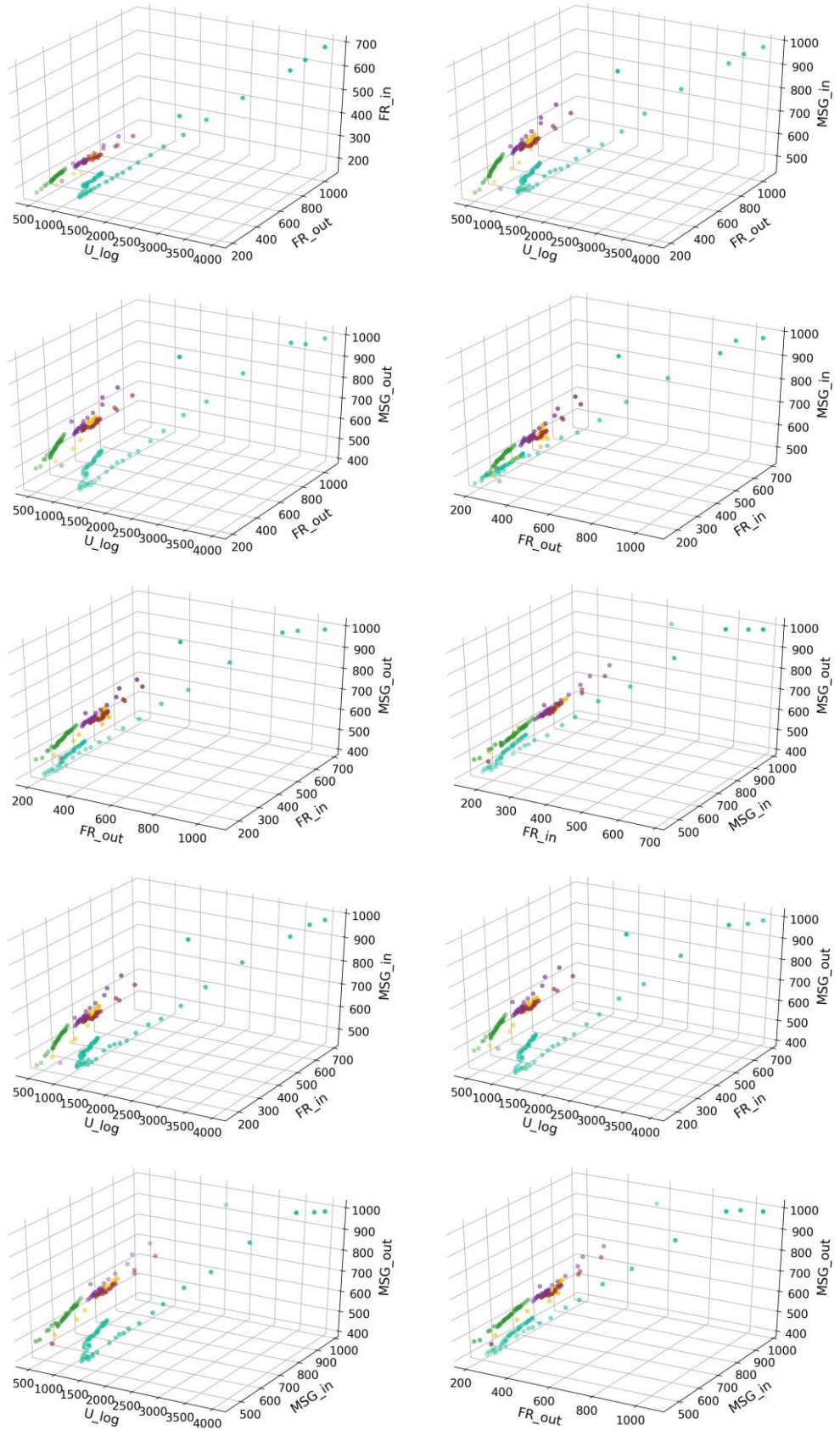

**Figure A2.** Dependence of objects in classes from all Experience Factors.